\documentclass[prd,nofootinbib,superscriptaddress]{revtex4}

\usepackage{amsmath}
\usepackage{amsfonts}

\begin{document}

\title{Probing the quantum-gravity realm with slow atoms}

\author{Flavio MERCATI}
\email{Spinoro@gmail.com}
\affiliation{Dipartimento di Fisica, Universit\`a di Roma ``La Sapienza"\\
and Sez.~Roma1 INFN, P.le A. Moro 2, 00185 Roma, Italy}

\author{Diego MAZ\'ON}
\affiliation{Departamento de F\'{\i}sica Te\'{o}rica, Facultad de Ciencias,\\
  Universidad de Zaragoza, 50009 Zaragoza, Spain}

\author{Giovanni AMELINO-CAMELIA}
\affiliation{Dipartimento di Fisica, Universit\`a di Roma ``La Sapienza"\\
and Sez.~Roma1 INFN, P.le A. Moro 2, 00185 Roma, Italy}

\author{Jos\'e Manuel CARMONA}
\affiliation{Departamento de F\'{\i}sica Te\'{o}rica, Facultad de Ciencias,\\
  Universidad de Zaragoza, 50009 Zaragoza, Spain}

\author{Jos\'e Luis CORT\'ES}
\affiliation{Departamento de F\'{\i}sica Te\'{o}rica, Facultad de Ciencias,\\
  Universidad de Zaragoza, 50009 Zaragoza, Spain}

\author{Javier INDUR\'AIN}
\affiliation{Departamento de F\'{\i}sica Te\'{o}rica, Facultad de Ciencias,\\
  Universidad de Zaragoza, 50009 Zaragoza, Spain}

\author{Claus L\"AMMERZAHL}
\affiliation{ZARM, Universit\"at Bremen, Am Fallturm, 28359 Bremen, Germany}

\author{Guglielmo M.~TINO}
\affiliation{Dipartimento di Fisica e
Astronomia and LENS, Universit\`a di Firenze,\\
Sez.~INFN di Firenze, Via Sansone 1, 50019 Sesto Fiorentino, Italy}


\begin{abstract}
For the study of Planck-scale modifications of the energy-momentum dispersion
relation, which had been previously focused
on the implications for ultrarelativistic (ultrafast) particles, we consider the
possible role of experiments involving nonrelativistic particles, and particularly atoms.
We extend a recent result establishing that measurements of  "atom-recoil frequency"
can provide insight that is valuable for some theoretical models.
And from a broader perspective we analyze the complementarity of the nonrelativistic
and the ultrarelativistic regimes in this research area.
\end{abstract}

\maketitle

\section{Introduction}
Over the last decade there has been growing interest in the
possibility to investigate experimentally some candidate
effects of quantum gravity.
The development of this ``quantum-gravity phenomenology"~\cite{gacLRR}
of course focuses on rare contexts in which the minute
effects induced by
ultra-high ``Planck
scale" $M_P ( \equiv \sqrt{ \hbar c / G} \simeq 1.2  \cdot 10^{28} ~ \text{eV})$
are not completely negligible.
Several contexts of this sort have been found
particularly
in the study of quantum-gravity/quantum-spacetime effects
for the propagation of ultrarelativistic/ultrafast particles
(see, {\it e.g.},
Refs.~\cite{grbgac,astroSchaefer,astroBiller,gacNature1999,urrutiaPRL,jaconature,Gaclaem,PiranNeutriNat,hessPRL}),
and often specifically for cases in which
the ultrarelativistic on-shell condition\footnote{We
adopt units in which the speed-of-light
scale $c$ is set to $1$ (whereas we shall explicitate the role of the
Planck constant $h$).}, $E \simeq p + m^2/(2p)$, is modified by Planck-scale effects.

In the recent Ref.~\cite{gacPRL2009} some of us observed that
experiments involving cold (slow, nonrelativistic) atoms,
and particularly measurements of the atom-recoil frequency,
can provide valuable insight
on certain types
of modifications of the dispersion relation which had been previously considered in
quantum-gravity literature.
We here extend the scopes of the analysis briefly reported in
Ref.~\cite{gacPRL2009}, also adopting a style of presentation that allows to
comment in more detail the derivation of the result. Concerning the
conceptual perspective that guides this recent research proposal,
we here expose some previously unnoticed aspects of complementarity between the nonrelativistic
and the ultrarelativistic regimes in the study of Planck-scale modifications
of the dispersion relation.
And we offer several observations on how the insight gained from studies
of slow atoms might translate into limits of different strength depending
on some details of the overall framework within which the modifications
of the dispersion relation are introduced.
We also report a preliminary exploration of the relativistic issues
involved in these studies, which have been already well appreciated in the
ultrarelativistic regime but appear to provide novel challenges
when the focus is instead on the nonrelativistic regime.

\section{Complementarity of nonrelativistic and ultrarelativistic regimes}\label{COMPLEMENTARITY}
Results in support of the possibility of modifications of the
energy/momentum (``dispersion")
relation have been reported in studies of several approaches to the
quantum-gravity problem,
and perhaps most notably in analyses inspired by
Loop Quantum Gravity~\cite{urrutiaPRL,LQGDispRel},
and in studies that assumed a ``noncommutativity"
of spacetime coordinates~\cite{gacmajid,kowaPLBcosmo,Orfeupion}.
The analyses of these quantum-gravity approaches
that provide encouragement for the presence of corrections to the dispersion
relation have become increasingly robust over the last
decade \cite{LQGDispRel,smolinbook,gacmajid,kowaPLBcosmo,Orfeupion},
but in the majority of cases
they are still unable to establish robustly the functional dependence of
the correction on
momentum. This has led to the proposal that perhaps on this occasion
experiments might take the lead by establishing
some experimental facts (at least in the form of constraints on the form
of the
dispersion relation) that may provide guidance for the ongoing
investigations on the theory side.

In light of these considerations the majority
of phenomenological studies of Planck-scale corrections to the
dispersion relation
have assumed
a general {\it ansatz},
\begin{eqnarray}
E^2 = p^2 + m^2 + \Delta_{QG}(p,m,M_P)~,
\label{generalansatz}
\end{eqnarray}
denoting with $E$ the energy of the particle
and with $\Delta_{QG}$ a model-dependent function of
the Planck mass $M_P$ and of the spatial momentum $p$
and of the mass $m$ of the particle.

Different models do give (more or less detailed) guidance on the form of
$\Delta_{QG}$,
and we shall consider this below,
but even at a model-independent level a few characteristics can be assumed
with reasonable robustness\footnote{We should stress however that,
while the perspective schematized in our Eqs.~(\ref{PropertiesOfDeltaQG})-(\ref{DispRelPrimoOrdineInMp})
is by far the most studied in the relevant quantum-gravity-inspired literature,
in principle more general possibilities may well deserve investigation. For example,
one might contemplate non-integer powers of $M_P$ to appear, and this would not be too
surprising, especially in light of the rather common expectation that
the correct description of quantum gravity might require sizable nonlocality.}.
As most authors in the field, we shall here focus our analysis
on cases in which the mass $m$ still is the rest-energy
and the dispersion
relation regains its ordinary special-relativistic form in the limit
where the
Planck scale is removed (${M_P} \rightarrow \infty$):
\begin{equation}
\Delta_{QG}(p,m,M_P) \xrightarrow[p \to 0]{} 0 ~, \qquad
\Delta_{QG}(p,m,M_P) \xrightarrow[M_P \to \infty]{} 0~.
\label{PropertiesOfDeltaQG}
\end{equation}
And these are most fruitfully exploited, since the relevant
phenomenology clearly can at best hope
to gain insight on the leading terms of a small-$M_P^{-1}$ expansion,
 within a power-series expansion,
\begin{equation}
E^2 = p^2 + m^2 + \frac{1}{M_P} \Delta_{QG}^{(1)}(p,m)
+ \frac{1}{M_P^2} \Delta_{QG}^{(2)}(p,m) + \dots ~,
\label{DispRelPrimoOrdineInMp}
\end{equation}
where the terms in the power series are subject to
the condition $\left. \Delta_{QG}^{(1)}(p,m)\right|_{p=0} =0 =\left.
\Delta_{QG}^{(2)}(p,m)\right|_{p=0}$.

This past decade of vigorous investigations of these modifications of
the dispersion
relation focused primarily (but not exclusively) on terms linear in
$M_P^{-1}$
and reached its most noteworthy results in analyses of observational
astrophysics data,
which of course concern the ``ultrarelativistic" ($p\gg m$) regime of
particle kinematics~\cite{grbgac,astroSchaefer,astroBiller,astroKifune,gacQM100,jaconature}.
For these applications
the function $\Delta_{QG}^{(1)}(p,m)$ can of course be usefully parametrized
in such a way that
the relation between energy
and spatial momentum takes the following form:
\begin{equation}
E \simeq p + \frac{m^2}{2p} + \frac{1}{2M_P}\left( \eta_1 \, p^2 +
\eta_2 \, m \, p + \eta_3 \, m^2 \right) ~,
\label{DispRelRelativistica}
\end{equation}
where, considering the large value of $M_P$, we
only included correction terms that are linear in $1/M_P$, and, considering that this formula
concerns the ultrarelativistic regime of $p \gg m$, the labels on the
parameters $\eta_1 , \eta_2 , \eta_3$ reflect the fact that in that regime $p^2/M_P$
is the leading correction,  $m p/M_P$ is next-to-leading, and so on.

Evidence that at least some of these $\eta_1$, $\eta_2$, $\eta_3$
parameters have nonzero values is indeed found in studies inspired by the
Loop-Quantum-Gravity approach
and by the approach based on spacetime noncommutativity, and most
importantly
some of these
studies~\cite{urrutiaPRL,LQGDispRel,gacmajid,kowaPLBcosmo,Orfeupion}
provide encouragement for the presence of the strongest imaginable
ultrarelativistic
correction, the leading-order term $\eta_1 \, p^2/(2M_P)$.

Unfortunately, as usual in quantum-gravity research,
even the most optimistic estimates represent a gigantic challenge from
the perspective
of phenomenology. This is because, if the Planck scale is indeed
roughly the
characteristic scale of quantum-gravity
effects then correspondingly parameters such as $\eta_1$, $\eta_2$, $\eta_3$
should take (positive or negative) values that are within no more than 1
or 2 orders
of magnitude of $1$.
And this in turn implies that, for example,
all effects induced by Eq.~(\ref{DispRelRelativistica}) could only
affect the running of our present particle-physics colliders
at the level~\cite{gacLRR} of at best 1 part in $10^{14}$.
In recent years certain semi-heuristic renormalization-group
arguments (see, {\it e.g.}, Refs.~\cite{gacLRR,wilcGUTEP} and
references therein),
have encouraged the intuition that the quantum-gravity scale might be
plausibly even 3 orders of
magnitude smaller than the Planck scale
(so that it could coincide~\cite{wilcGUTEP} with
the ``grand unification scale"
which appears to play a role in particle physics).
But even assuming for $\eta_1$, $\eta_2$, $\eta_3$
values plausibly as ``high" as $10^3$
is not enough help at traditional high-energy particle-collider experiments.

It was therefore rather exciting for many quantum-gravity researchers
when it started to emerge that some observations in astrophysics could
be sensitive to
manifestations of the parameter $\eta_1$ all the way down to $|\eta_1|
\sim 1$
and even
below~\cite{grbgac,astroSchaefer,astroBiller,astroKifune,gacQM100,jaconature},
thereby providing for that parameter the ability to explore the full
range of
values that could be motivated from a quantum-gravity perspective.
These studies are presently being conducted at the
Fermi Space
Telescope~\cite{fermiSCIENCE,ellisPLB2009,gacSMOLINprd2009,fermiNATURE,gacNATURE2009},
and other astrophysics observatories.

In the recent Ref.~\cite{gacPRL2009} some of us observed that it would
be very valuable
to combine to these astrophysics studies of the ultrarelativistic regime
of the dispersion
relation also a complementary phenomenology program of investigation
of the
nonrelativistic regime.
And in the regime of $p \ll m$
the 3 largest contributions to $\Delta_{QG}^{(1)}(p,m)$
have behavior\footnote{Note that a contribution
of form $m^3$ ({\it i.e.} momentum-independent) to
$\Delta_{QG}^{(1)}(p,m)$ cannot be included
in the nonrelativistic regime because of the requirement $\Delta_{QG}^{(1)}(p=0,m)  =0$.
A contribution to $\Delta_{QG}^{(1)}(p,m)$
of form $m^3$ is instead admissible in the ultrarelativistic regime
(since in that regime the requirement $\Delta_{QG}^{(1)}(p=0,m) =0$
of course is not relevant), but we ignored it since $m^3$ is too
small with respect
to $p^3$, $m p^2$ and $m^2 p$ in
the ultrarelativistic regime.}
 $m^2 p$, $m p^2$ and $p^3$,
allowing to cast the relation between energy and spatial momentum
in the following form:
\begin{equation}
E \simeq m + \frac{p^2}{2m} + \frac{1}{2M_P}\left( \xi_1 m p + \xi_2 p^2
+ \xi_3 \frac{p^3}{m} \right) \label{DispRelNonRelativistica}~,
\end{equation}
where, again, $\xi_1$, $\xi_2$, $\xi_3$ are dimensionless
parameters.

Evidence that at least some of these
dimensionless parameters $\xi_1$, $\xi_2$, $\xi_3$
should be non-zero has been found for example
in the much-studied framework introduced in
Refs.~\cite{urrutiaPRL,urrutiaPRD},
which was inspired by Loop Quantum Gravity,
and produces a term linear in $p$
in the nonrelativistic limit (the effect here parametrized by $\xi_1$).
And for the purposes of this Section, which
we are devoting to the complementarity of the
nonrelativistic and ultrarelativistic regimes of the dispersion relation,
it is particularly insightful to consider
two of the most studied scenarios
that have emerged in the literature on noncommutative-geometry-inspired
deformations of Poincar\'e symmetries.
These are the scenarios proposed in
Refs.~\cite{gacIJMP2002vD11,gacdsrPLB2001}
and in Ref.~\cite{leedsrPRL}, which respectively produce the following
proposals
for the exact form of the dispersion relation:
\begin{equation}
\left(\frac{2 M_P}{\eta}\right)^2 \sinh^2 \left(\frac{\eta E}{2 M_P}
\right) = \left(\frac{2 M_P}{\eta}\right)^2 \sinh^2 \left(\frac{\eta
m}{2 M_P} \right) + e^{-\eta \frac{E}{M_P}} p^2~, \label{DSRs1}
\end{equation}
and
\begin{equation}
\frac{m^2}{(1- \eta\frac{m}{M_P} )^2} =
\frac{E^2-p^2}{(1-\eta\frac{E}{M_P})^2} ~, \label{DSRs2}
\end{equation}
Both of these proposals have the same description in the nonrelativistic
regime
\begin{equation}
E \simeq m + \frac{p^2}{2m} - \eta \frac{p^2}{2 M_P} ~,
\end{equation}
{\it i.e.} the type of correction term
in the nonrelativistic regime that we are here parameterizing with $\xi_2$.
But these proposals
have significantly different behavior in the ultrarelativistic regime.
 From Eq.~(\ref{DSRs1}) in
the ultra-relativistic regime one finds
\begin{equation}
E \simeq p + \frac{m^2}{2p} - \eta \frac{p^2}{2M_P} ~,
\end{equation}
whereas from Eq.~(\ref{DSRs2}) in
the ultra-relativistic regime one finds
\begin{equation}
E \simeq p + \frac{m^2}{2p} - \eta \frac{m^2}{M_P} ~.
\end{equation}
Therefore the example of these two much studied deformed-symmetry
proposals is such that by focusing exclusively on the nonrelativistic
regime one
could not (not at the leading order at least) distinguish between them,
but one could discriminate between the two proposals using data on the
ultrarelativistic
regime.
The opposite is of course also possible: different candidate dispersion
relations
with the same ultrarelativistic limit, but with different leading-order
form in
the nonrelativistic regime.
And in general it would be clearly very valuable to constrain the form
of the
dispersion relation both using experimental information on the leading
nonrelativistic behavior and
experimental information on the leading ultrarelativistic behavior.

\section{Probing the nonrelativistic regime with cold atoms}\label{nonrUR}
Our main objective here is to show
that cold-atom experiments
can be valuable for the study of Planck-scale effects.
We illustrate this point mainly by considering
the possibility, already preliminarily
characterized in Ref.~\cite{gacPRL2009},
to use cold-atom studies for the derivation of
 meaningful bounds on the parameters $\xi_1$
and $\xi_2$, {\it i.e.} the leading and next-to-leading terms
in (\ref{DispRelNonRelativistica}) for the nonrelativistic limit:
\begin{equation}
E \simeq m + \frac{p^2}{2m} + \frac{1}{2M_P}\left( \xi_1 m p + \xi_2 p^2  \right) ~.
\label{DispRelNonRelativisticaJOC}
\end{equation}
In this section we work exclusively from a laboratory-frame
perspective, as done in Ref.~\cite{gacPRL2009},
but, as for most relativistic studies, it is valuable to also perform the analysis
in one or more frames that are boosted with respect to the laboratory frame,
and we shall discuss this
in Sec.~\ref{BOOSTEDFRAME}.

The measurement strategy proposed  in Ref.~\cite{gacPRL2009} is applicable to measurements
of the ``recoil frequency" of atoms with experimental setups
involving one or more ``two-photon Raman transitions"~\cite{Kasevich91b,Peters99,Wicht02}.
Let us initially set aside the possibility of Planck-scale effects,
and discuss
the recoil of an atom in a two-photon Raman transition
from the perspective adopted
in Ref.~\cite{Wicht02},
which provides a convenient starting point for the Planck-scale generalization
we shall discuss later.
One can impart momentum to an atom through a process involving absorption
of a photon of frequency $\nu$ and (stimulated~\cite{Kasevich91b,Peters99,Wicht02})
emission, in the opposite direction,
of a photon of frequency $\nu'$. The frequency $\nu$ is computed taking into
account a resonance frequency $\nu_*$
of the atom and the momentum the atom acquires,  recoiling upon absorption
of the photon: $ \nu \simeq  \nu_* + ( h \nu_* + p)^2/(2 m) - p^2/(2m)$,
where $m$ is the mass of the atom ({\it e.g.} $m_{Cs} \simeq 124~\text{GeV}$ for Caesium),
and $p$ its initial momentum.
The emission of the photon of frequency $\nu'$ must be such to
de-excite\footnote{We only give a schematic and simplified account of the process,
 which suffices for the
scopes of our analysis. A more careful description requires taking into account
that, rather than a single ground state, the relevant two-photon Raman transition
involve hyperfine-splitted ground states~\cite{Kasevich91b,Peters99,Wicht02}. And that, rather than
tuning the two lasers exactly on some energy differences
between levels, some detuning is
 needed~\cite{Kasevich91b,Peters99,Wicht02}.}
  the atom and impart to it additional
momentum: $\nu' + (2 h \nu_* + p)^2/(2 m) \simeq  \nu_*
  + (h \nu_*+p)^2/(2 m)$.
Through this analysis one establishes that by measuring $\Delta \nu \equiv \nu - \nu'$,
in cases (not uncommon) where $\nu_*$ and $p$ can be accurately determined, one
actually measures $h/m$ for the atoms:
\begin{eqnarray}
     \frac{\Delta \nu}{ 2 \nu_* (\nu_* +p/h)} = \frac{h}{m} ~. \label{deltaomeNOEP}
\end{eqnarray}
This result has been confirmed experimentally with remarkable accuracy.
A powerful way to illustrate this success is provided by comparing
the results for atom-recoil
measurements of $\Delta \nu/[\nu_* (\nu_* +p/h)]$ and for measurements~\cite{gab08}
of $\alpha^2$, the square
of the fine structure constant. $\alpha^2$ can be expressed in terms of the mass $m$
of any given particle~\cite{Wicht02} through the Rydberg constant, $R_\infty$,
 and the mass of the electron, $m_{{e}}$, in the following
 way~\cite{Wicht02}: $ \alpha^2 = 2 R_\infty \frac{m}{m_{{e}}} \frac{h}{m}$.
Therefore according to Eq.~(\ref{deltaomeNOEP}) one should have
\begin{equation}
\frac{\Delta \nu}{ 2 \nu_* (\nu_* +p/h)} =
\frac{\alpha^2}{2 R_\infty}
\frac{m_e}{m_u} \frac{ m_u}{m} ~, \label{alphaJ2}
\end{equation}
where $m_u$ is the
atomic mass unit and $m$ is the mass of the atoms used in
measuring $\Delta \nu/[\nu_* (\nu_* +p/h)]$. The outcomes of atom-recoil measurements,
such as the ones with Caesium reported in Ref.~\cite{Wicht02},
are consistent with Eq.~(\ref{alphaJ2})
with the accuracy
of a few parts in $10^9$.

The fact that Eq.~(\ref{deltaomeNOEP}) has been verified to such a high degree
of accuracy proves to be very
valuable for our purposes as we find
that modifications of the dispersion relation require a modification
of  Eq.~(\ref{deltaomeNOEP}). Our derivation can be summarized briefly
by observing that
the logical steps described above for the derivation of Eq.~(\ref{deltaomeNOEP})
establish the following relationship
\begin{equation}
 h \Delta \nu \simeq  E(p + h\nu + h\nu') - E(p) \simeq E(2  h\nu_* + p) - E(p) ~, \label{DeltaOmegaGenerico}
\end{equation}
and therefore  Planck-scale modifications of the
dispersion relation,
parametrized in Eq.~(\ref{DispRelNonRelativistica}),
would affect $\Delta \nu$ through the modification
of $E(2 h \nu_* + p) - E(p)$, which compares the energy of
the atom when it carries momentum $p$ and when it carries momentum $p+2 h \nu_*$.

Since our main objective here is to expose sensitivity to a meaningful range of
values of the parameter $\xi_1$,
let us focus on the Planck-scale corrections
 with coefficient $\xi_1$.
In this case the relation (\ref{deltaomeNOEP}) is replaced by
\begin{eqnarray}
\Delta \nu \!  & \simeq & \! \frac{ 2 \nu_* (h \nu_* +p)}{m} +  \xi_1 \frac{m}{M_P} \nu_*
~,
\label{DeltaOmegaLeading}
\end{eqnarray}
and in turn in place of Eq.~(\ref{alphaJ2})
one has
\begin{equation}
\frac{\Delta \nu}{ 2 \nu_* (\nu_* \! + \! p/h)} \!\! \left[ \! 1
\! - \xi_1 \! \left( \! \frac{ m}{2 M_P} \! \right) \!\! \left( \! \frac{m}{h \nu_* +p} \! \right)
 \! \right] \!\! = \!\!
\frac{\alpha^2}{2 R_\infty}
\frac{m_e}{m_u} \frac{ m_u}{m} ~.
\label{jocMINUS1}
\end{equation}
We have arranged the left-hand side of this equation placing emphasis on the
fact that our quantum-gravity
correction is as usual penalized
 by the inevitable Planck-scale suppression (the ultrasmall factor $m /M_P$),
 but in this specific context it also receives a sizable boost by the large
 hierarchy of energy scales $m/(h \nu_* +p)$, which
in typical experiments of the type here of interest can be~\cite{Kasevich91b,Peters99,Wicht02}
of order $\sim 10^{9}$.

Our result (\ref{jocMINUS1}) for the case of modification of the dispersion relation
by the term with coefficient $\xi_1$
can be straightforwardly generalized
to the case of a modified dispersion relation of the form
\begin{equation}
E \simeq m + \frac{p^2}{2m} +
\frac{\xi_{\beta}}{2}
\frac{m^{2-\beta}}{M_P} p^{\beta}
 \label{DispRelJOSE}
\end{equation}
which reproduces our terms with parameters $\xi_1$ and $\xi_2$ respectively
when $\beta =1$ and $\beta = 2$ (but in principle could be examined even
for non-integer values of $\beta$).

One then finds
\begin{equation}
\frac{\Delta \nu}{2\nu_*(\nu_*+p/h)}\left[1 - \xi_{\beta}
\left(\frac{m^{2-\beta}\left[(p+2h\nu_*)^{\beta} -
    p^{\beta}\right]}{4M_P h\nu_*}\right)
\left(\frac{m}{h\nu_*+p}\right)\right] \,=\, \frac{\alpha^2}{2
  R_{\infty}} \frac{m_e}{m_u} \frac{m_u}{m}
\end{equation}
which indeed reproduces (\ref{jocMINUS1}) for $\beta = 1$ and gives~\cite{gacPRL2009}
\begin{equation}
\frac{\Delta \nu}{ 2 \nu_* (\nu_* \! + \! p/h)} \!\! \left[  1
\! - \xi_2 \frac{m}{M_P}
 \right] \!\! = \!\!
\frac{\alpha^2}{2 R_\infty}
\frac{m_e}{m_u} \frac{ m_u}{m} ~, \label{jocMINUS2}
\end{equation}
for $\beta = 2$.

\section{Limits on different models}\label{LIMITS}
From a phenomenological perspective
the most
remarkable observation one can ground on the results reported in the previous Section
is that the accuracies achievable in cold-atom studies
allow us to probe values of $\xi_1$ that are not distant from $|\xi_1| \sim 1$.
This is rather meaningful since, as stressed in the previous Section,
the quantum-gravity intuition for parameters such as $\xi_1$ is that
 they should be (in models where a nonzero value for them is allowed)
 within a few orders of magnitude of 1.
Besides discussing this point, in this Section we also consider the case of the term
with $\xi_2$ parameter and we comment on the relevance of these analyses
from the perspective of a class of phenomenological proposals which is broader
than the one here discussed in Section II.
The closing remarks of this Section are devoted to
observations that may be relevant for attempts to further improve the
relevant experimental limits.

\subsection{Limits on $\xi_1$ and $\xi_2$}
The fact that our analysis provides sensitivity to values of $\xi_1$ of order 1
is easily verified by examining our result for the case of the $\xi_1$ parameter,
which we rewrite here for convenience
\begin{equation}
\frac{\Delta \nu}{ 2 \nu_* (\nu_* \! + \! p/h)} \!\! \left[ \! 1
\! - \xi_1 \! \left( \! \frac{ m}{2 M_P} \! \right) \!\! \left( \! \frac{m}{h \nu_* +p} \! \right)
 \! \right] \!\! = \!\!
\frac{\alpha^2}{2 R_\infty}
\frac{m_e}{m_u} \frac{ m_u}{m} ~,
\label{joclong1}
\end{equation}
and taking into account some known experimental accuracies.
Let us focus in particular on the Caesium-atom recoil measurements
reported in Ref.~\cite{Wicht02}, which were ideally structured for our purposes.
Let us first notice that $R_\infty$, ${m_e}/{m_u}$ and ${ m_u}/{m_{Cs}}$
are all known experimentally with accuracies of better than 1 part in $10^9$.
When this is exploited in combination with the value
of ${\alpha^2}$ recently determined from electron-anomaly measurements~\cite{gab08},
which is
 ${\alpha^2} = 137.035 999 084 (51)$, the results of  Ref.~\cite{Wicht02,Gerginov06}
then allow us to use (\ref{joclong1}) to determine that $\xi_1 = - 1.8 \pm 2.1$.
This amounts to
the bound $-6.0  < \xi_1 < 2.4 $,
 established at the 95\% confidence level,
and shows that indeed the cold-atom experiments we here considered can probe the form of the dispersion
relation (at least in one of the directions of interest) with sensitivity that
is meaningful from a Planck-scale perspective.

As mentioned in Section~\ref{COMPLEMENTARITY} among the models that could be here of interest
there are some where, by construction, $\xi_1=0$ but $\xi_2 \neq 0$.
In such cases it is then of interest to establish bounds on $\xi_2$ derived
assuming $\xi_1=0$, for which one can easily adapt the derivation discussed above.
These are therefore cases in which our result (\ref{jocMINUS2}) is relevant,
and one easily then finds that the atom-recoil results for
Caesium atoms
reported in Refs.~\cite{Wicht02,Gerginov06} can be used
to establish that  $- 3.8 \cdot 10^{9} < \xi_2 < 1.5 \cdot 10^{9}$.
This bound is
 still some 6 orders of magnitude above even the most optimistic
 quantum-gravity estimates.
But it is a bound that still carries some significance
from the broader perspective of tests of Lorentz symmetry~\cite{gacPRL2009}.

\subsection{Relevance for other quantum-gravity-inspired scenarios}
Up to this point we have assumed ``universal" effects,
{\it i.e.} modifications of the dispersion relation that have the same
form for all particles, independently of spin and compositeness,
and with dependence on the mass of the particles rigidly inspired by
the quantum-gravity arguments suggesting correction terms of the
form $m^j p^k/M_{p}^l$
({\it i.e.} with a characteristic dependence on momentum and with a momentum-independent
coefficient written as a ratio of some power of the mass of the particle versus
some power of the Planck-scale).

While this universality is indeed assumed in the majority of studies of
the fate of Poincar\'e symmetry at the Planck scale,
alternatives have been considered by some authors~\cite{liberatiNONUNIV}
and there are good reasons to at least be open to the possibility of nonuniversality.
One reason of concern toward universality originates from the fact that clearly
modifications of the dispersion relation at the Planck scale
are a small effect for microscopic particles (always with energies much below the Planck
scale in our experiments), but would be a huge (and unobserved) effect for macroscopic
bodies, such as planets and, say, soccer balls.
Even the literature that assumes universality is well aware of this issue, and in fact
the opening remarks of papers on this subject always specify a restriction
to microscopic particles. With our present (so limited) understanding of the quantum-gravity
realm we can indeed contemplate for example the possibility that such effect
be confined to motions which admit description in terms of coherent quantum systems
(by which we simply mean that the focus is on the type of particles whose quantum properties
could also be studied in the relevant class of phenomena, unlike the motions of
planets and soccer balls).
This is clearly (at least at present) a plausible scenario that many authors are studying
and for which atoms provide an extraordinary opportunity of investigation of
the nonrelativistic limit: because of their relatively large masses atoms have ultrashort
(de Broglie) wavelengths even at low speeds and provide relatively large values for
terms of the form $m^j p^k/M_{p}^l$. Let us compare for example our study to the popular studies
of the ultrarelativistic regime with photons. The best limits on the ultrarelativistic side
are obtained~\cite{fermiNATURE} through observations of photons with energies of a few
tens of GeV's. The limit we here obtained in the nonrelativistic regime involves very small
speeds ($\ll c$) but for particles, the atoms, with (rest) energies in the $\sim 100~\text{GeV}$
range.

While it is therefore rather clear that atoms are excellent probes of scenarios
with universality for ``quantum-mechanically microscopic particles", their effectiveness
can be sharply reduced in models with some forms of nonuniversality.
In particular, one could consider the compositeness of particles
as a possible source of nonuniversality~\cite{dsrIJMPrev}.
And this would imply that in the study of processes involving, say, protons
and pions one should adopt a ``parton picture" with the number of partons
acting in the direction of averaging out the effects: if quantum-spacetime
effects affect primarily the partons then a particle composed of 3 partons
could feel the net result of 3 such fundamental features, with a possible
suppression ({\it e.g.} by a factor of $\sqrt{3}$)
of the effect for the particle with respect to the fundamental effect for partons.
These ideas have not gained much attention, probably also because things might change only
at the level of factors of order $1$ if one was for example to devise ways to keep track
of the different number of partons for nucleons and for pions.
But in the case of atoms, that we are now bringing to the forefront of quantum-gravity phenomenology,
clearly these concerns cannot be taken lightly: for the description of an atom one
might have to consider hundreds of partons (or at least $\sim 100$ nucleons).
We therefore expect that our strategy to place limits on $\xi_1$ and $\xi_2$
will be less effective (limits more distant from the Planck scale) in scenarios based
on one or another form of ``parton model" for the implications of spacetime quantization
on quantum-mechanical particles.
We do not dwell much on this here at the quantitative level since the literature
does not offer us definite models of this sort that we could compare to data.

Even assuming that the effect is essentially universal one could
consider alternatives to the most common assumption that
quantum-gravity corrections have the form $m^j p^k/M_P^l$.
In particular, some authors (see, {\it e.g.}, Refs.~\cite{HIN,HOS,ROV})
have argued that the density of energy (or mass) of a given particle
(be it elementary or composite) should govern the magnitude of the effect,
rather than simply the mass of the particle.
This is another possibility which is also under investigation~\cite{HIN,HOS,ROV}
as a mechanism
for effectively confining the new effects to elementary particles.
In the simplest scenarios this proposal might amount to
replacing terms such as our $\xi_1 m p/(2M_P)$
with terms of the general
form ${\tilde{\xi}}_1 \rho^{1/4} p/(2 M_P)$,
but of course the implications of such pictures depend crucially
on exactly which density $\rho$ one adopts.
For different choices of $\rho$ the limits  derived from atom-recoil experiments
can be more or less stringent than
those derived in studies of lighter particles,
such as electrons.

Another framework which can be used to illustrate the different
weight that cold-atom studies can carry in different scenarios
for the deformation of the dispersion relation is the
one already studied in Refs.~\cite{LIV-tritium,LIV-cutoffs},
parameterized by a single scale $\lambda$ such
that $E^2=m^2+p^2+2\lambda p$.
Limits on this form of the dispersion
relation have been obtained for neutrinos in Ref.~\cite{LIV-tritium},
and for electrons, in Ref.~\cite{LIV-cutoffs}. Taking into account
that from $E^2=m^2+p^2+2\lambda p$ it follows that in the
nonrelativistic limit $E=m+p^2/(2m)+\lambda p/m$, one easily finds
that the parametrization we introduced in Eq.~(\ref{DispRelNonRelativistica})
and the parametrization of
Refs.~\cite{LIV-tritium,LIV-cutoffs} are related by $\xi_1 m/M_P\equiv 2 \lambda/m$.
And in light of this one easily sees that our atom-recoil analysis
can also be used to establish the
bound $-3.7\cdot 10^{-6}\,\text{eV}<\lambda<1.5\cdot 10^{-6}\,\text{eV}$.
This shows
that the cold-atom-based strategy is suitable also for studies
of the $\lambda$-parameter picture of Refs.~\cite{LIV-tritium,LIV-cutoffs}.
But, while, as some of us already stressed in Ref.~\cite{gacPRL2009},
this atom-based bound on $\lambda$ is more powerful (by roughly 6 orders of magnitude)
than bounds previously obtained on $\lambda$ using neutrino data~\cite{LIV-tritium},
we should here notice that the best present bound on $\lambda$ is the electron-based
bound derived in Ref.~\cite{LIV-cutoffs}, which is at the level $|\lambda| \lesssim 10^{-7}~\text{eV}$.
We stress that there is no contradiction between the remarks we offered above on the unique
opportunities that cold-atom studies provide for setting bounds on the parameter $\xi_1$,
and the fact that instead for the $\lambda$ parameter
electron studies are competitive with
(and actually still slightly more powerful than)  atom-based studies:
this difference between the strategies for bounding the $\xi_1$ parameter and the $\lambda$
parameter is easily understood in light of the relation  $\xi_1 m/M_P \leftrightarrow 2 \lambda/m$
and of the large difference of masses between electrons and (Caesium or Rubidium) atoms.

Finally, in closing this Subsection on alternative models,
let us mention the possibility of intrinsically non-universal modifications
of the dispersion relation, {\it i.e.} phenomenological scenarios in which
the modifications of the dispersion relation are assumed to be different
for different particles without introducing any specific prescription
linking these difference to the mass, the spin or other specific
properties of the particles. For example, in Ref.~\cite{liberatiNONUNIV},
and references therein, the authors introduce a free parameter for each different type of
particle. In such cases studies of Caesium and, say, Rubidium atoms could be used to
set constraints on parameters that are specialized to those types of atoms. In essence,
according to this (certainly legitimate) perspective, we might learn that for Caesium
and Rubidium $\xi_1$ is small but without assuming any implications for the values
of $\xi_1$ for other particles.
And another noteworthy example is the one of Ref.~\cite{ellisPHOTONonly}, and references
therein, where it is argued, within a specific scenario for quantum gravity, that the effects
of modification of the dispersion relation should be confined to a single type
of particle, the photon (in which case of course atoms cannot possibly be of any help).

\subsection{Strategies for improving the limits}\label{NEWLIMIT}
As a contribution toward the development of experimental setups which in some cases
may be optimized for our proposal it is important for us to stress that, while essentially
here we structured our analysis in a way that might appear to invite interpretation
as ``quantum-gravity
corrections to $h/m$ measurements", not all improvements in the sensitivity of measurements
of $h/m$ will translate into improved bounds on
the parameters we here considered.

First we should notice that our result for the $\xi_1$-dependent correction
to $\Delta \nu /[ 2 \nu_* (\nu_* \! + \! p/h)]$ would not appear as a constant shift
of $h/m$, identically applicable to all experimental setups.
This is primarily due to the fact that, as shown in Eq.~(\ref{joclong1}),
our quantum-gravity correction factor has the
form $ 1 - \xi_1  m^2/[2 M_P(h \nu_* +p)]$, and therefore at the very least should be viewed
as a momentum-dependent shift of $h/m$. Different $h/m$ measurements,  even when relying on the
same atoms (same $m$), are predicted to find different levels of inconsistency with the
uncorrected relationship between $h/m$ and $\alpha^2$.
This is particularly important because some of the standard techniques~\cite{Wicht02,udem} used
to improve the accuracy of measurement of $h/m$ rely on imparting to the atoms
higher overall values of momentum, but since the magnitude of the $\xi_1$-governed effect
decreases with the magnitude of momentum, these possible ways to
get more accurate determinations of $h/m$ might not actually provide more stringent bounds
on $\xi_1$.
This is after all one of the reasons why the bound on $\xi_1$ which we discussed
here relied on the determinations of $h/m$ reported in Ref.~\cite{Wicht02,Gerginov06}:
a more accurate determination of $h/m$ was actually obtained in
the cold-atom (Rubidium) studies reported in Refs.~\cite{biraben06,biraben08},
but those more accurate determinations of $h/m$ relied on much higher
values of momentum, thereby producing a bound on $\xi_1$ which is not competitive~\cite{gacPRL2009}
with the one obtainable using the $h/m$ determination of Ref.~\cite{Wicht02,Gerginov06}.
The challenge we propose is therefore the one of reaching higher accuracies
in the measurement of $h/m$ without increasing significantly the momentum
imparted to the atoms.

Interestingly these concerns do not apply to our result for the $\xi_2$ parameter.
In fact,
our result for the $\xi_2$-dependent correction
to $\Delta \nu /[ 2 \nu_* (\nu_* \! + \! p/h)]$ would actually appear as a constant shift
of $h/m$, a mismatch between $h/m$ results and $\alpha^2$ results
of identical magnitude in all experimental setups using the same atoms (same $m$).
This is due to the fact that, as shown in Eq.~(\ref{jocMINUS2}),
our quantum-gravity correction factor has the
form $[ 1 - \xi_2  m/M_P]$, and therefore can indeed be viewed
as a (mass-dependent but) momentum-independent shift of $h/m$.

Besides these issues connected with the role played by the momentum of the atoms
in our analysis, there are clearly other issues that should be taken into consideration
by colleagues possibly contemplating measurements of $h/m$ that could improve
the limits on our parameters. One of these clearly deserves mention here, and concerns
the setup of $h/m$ measurements as differential measurements.
In this respect it is rather significant that our derivation of
dependence of the measured $\Delta \nu$ on the Planck-scale effects
shows that the sign of the correction term depends on
the ``histories" (beam-splitting/beam-recombination histories)
of the atoms whose interference is eventually measured.
Even from this perspective our result is therefore not to be viewed
simply as ``a shift in $h/m$": often in the relevant cold-atom experiments
one achieves a very accurate determination of $h/m$ by comparing
(in the sense of a differential measurement)
two different values of $\Delta \nu$ obtained by interference of different pairs
of beams produced in the beam-splitting/beam-recombination sequence of
a given experimental setup.
We should therefore warn our readers
that for some differential measurements the effect measured would be twice
as large as the one we here computed (same effect
but with opposite sign on the two sides of the differential measurement),
but on the other hand it is not hard to
arrange\footnote{The careful reader will for example notice that
Ref.~\cite{birabenPROCEEDINGS} provides an example of setup in which
our Planck-scale effects would cancel out.} for a differential measurement
that is insensitive to the quantum-gravity effects (if the ``histories" are such that
the correction carries the same
sign on the two sides of the differential measurement).

\section{Atom velocity, energy-momentum conservation and other relativistic issues}\label{BOOSTEDFRAME}
We have so far focused
on schemes which assume that the only new relevant quantum-gravity-induced law
amounts to a modification of the energy-momentum dispersion relation.
The main results here derived in Section~\ref{nonrUR}
relied on a strategy of analysis
that only requires a specification in the ``laboratory frame"
of the form of the dispersion relation
(which is used to establish, for example, the energy gained by an atom when
its spatial momentum is increased)
and the law of energy-momentum conservation
(which is used to establish, for example, the spatial momentum imparted to an atom
upon absorption of a photon of known wavelength).
Even within that scheme of analysis one clearly should consider also the possibility
of modifications of the law of energy-momentum conservation, especially in light of the
fact that certain quantum-gravity scenarios establish (see below) a direct link between
modifications of the dispersion relation and some corresponding modifications
of the law of energy-momentum conservation.

Moreover, the laboratory-frame perspective is of course too narrow for the investigation
of the relativistic issues that clearly must be involved in scenarios
that introduce modifications of the dispersion relation. Also from this perspective
the quantum-gravity literature offers significant motivation for a careful investigation,
since modifications of the laws of transformation between reference frames have been
very actively studied (see below). And, as we shall here stress, connected to this issue
of boost transformations between reference frames one also finds intriguing
challenges for what concerns the description of the velocity of particles.

In this Section we offer an exploratory discussion of these issues.
Even in the quantum-gravity literature on ultrarelativistic modifications
of the dispersion relation the study of these issues has proven very challenging,
and many unsolved puzzles remain. So we shall not even attempt here to address fully
these issues in the novel domain of the nonrelativistic limit, which we are here advocating.
But we hope that the observations we report here may provide a valuable starting point
for more detailed future studies.

Among the ``exploratory aspects" of our discussion we should in particular stress
that we assume here, as done in most of the related quantum-gravity-inspired literature,
that concepts such as energy, spatial momentum and velocity can still be discussed
in standard way, so that the novelty of the pictures resides in new laws linking
symbols that admit a conventional/traditional physical interpretation.
Of course, alternative possibilities also deserve investigation:
a given quantum-gravity/quantum-spacetime picture might well (when fully understood)
provide motivation not only for novel forms of, say, the dispersion relation but also
impose upon us a novel description of the entities that appear in the dispersion
relation, such as a novel understanding of the energy $E$ that appears in the dispersion
relation. But we shall already highlight several challenges for the more conservative
scenario (with traditional ``interpretation of symbols"), and therefore
we postpone to future works the investigation of alternative interpretations.

\subsection{Velocity and boosted-frame analysis}
As a partial remedy to the laboratory-frame limitation
of the strategy of analysis discussed in Section~\ref{nonrUR},
we take as our next task the one of deriving the same result
using a scheme of derivation involving boosting and the Doppler effect.
The role played by transformation laws between different observer-frames
motivates part of our interest for
 this calculation, since investigations of the
fate of Poincar\'e symmetry in models with Planck-scale modifications of the
dispersion relation
must in general address the issue of whether the symmetries are ``deformed",
in the sense of the ``Doubly Special Relativity" scenario~\cite{gacIJMP2002vD11,gacdsrPLB2001},
or simply ``broken".
When the symmetry transformations are correspondingly ``deformed" the dispersion relation
will be exactly the same for all observers~\cite{gacIJMP2002vD11,gacdsrPLB2001}.
In the symmetry-breaking alternative scenario the laws of boosting are unmodified
and as a result one typically finds that the chosen form of the dispersion
relation only holds for one class of observers (at the very least one must expect~\cite{gacflavtsviuri}
observer dependence of the parameters
that characterize the modification of the dispersion relation).
And another aspect of interest for such analyses originates from the fact that
the description of the Doppler effect requires a corresponding description of
the velocity of the atoms, and therefore requires a specification
of the law that fixes the dependence of speeds on momentum/energy at the Planck scale:
this too is a debated issue,
with many authors favoring $v(p) = \partial E/\partial p$, but some support in
the literature also for some alternatives, the most popular of which is $v=p/E$.

As stressed in the opening remarks of this Section,
we are just aiming for a first exploratory characterization of these issues
and their possible relevance for our atom-recoil studies.
Consistently with these scopes
we assume that the Doppler effect (boosting) is undeformed and
that the dispersion relation
is an invariant law. This of course is only one (and a particularly peculiar) example
of combination of the possible formulations of the main issues here at stake,
but it suffices for exposing the potentially strong implications
that the choice of these formulations can have for the analysis.

Let us start by reanalyzing the recoil of atoms in terms of a Doppler effect,
neglecting initially the possible Planck-scale effects (which we shall reintroduce
later in this Section).
When an  atom absorbs a photon whose frequency is  $\nu$ in the laboratory frame,
in the rest frame of the atom the photon has frequency $\tilde \nu = \nu ( 1 - v)$, where $v$
is the speed of the atom in the lab frame (and for definiteness we are considering the case
of photon velocity parallel to the atom velocity).
Then in the rest frame, if
the absorption of the photon takes the atom to
 an energy level $h \nu_*$,
energy conservation takes the form
\begin{equation}
\tilde \nu \simeq \nu_* + \frac{h \nu_*^2}{2m}~,
\label{jocflav}
\end{equation}
which of course can also be equivalently rewritten in terms of the
lab-frame frequency of the photon
\begin{equation}
\nu \simeq \nu_* (1 + v) + \frac{h \nu_*^2}{2m}~,
\end{equation}
also neglecting a contribution of order $v \, {h \nu_*^2}/{m}$
which is indeed negligible in the nonrelativistic ($v \ll 1$) regime.

This photon absorption also takes the atom from velocity $v$ to velocity $v'$,
\begin{equation}
v' \simeq v + h \nu_* / m ~,
\end{equation}
in the laboratory frame (where we also observed that the gain of momentum
of the atom is approximately $h\nu_*$).

For the stage of (stimulated) emission of a second photon, whose frequency in the lab frame
we denote with $\nu'$, the atom would then
be moving at this speed $v'$, and in the rest frame of the atom the frequency of
this emitted photon is $\tilde \nu' = \nu' (1+v')$ (also taking into account that if, in the lab frame,
 the absorbed photon
moved in parallel with the atom, the emitted photon must then move in anti-parallel direction).
In the case of photon emission, conservation of energy
in the rest frame has a different sign with respect to Eq.~(\ref{jocflav}), {\it i.e.}
\begin{equation}
\tilde \nu' \simeq \nu_* - \frac{h \nu_*^2}{2m}~,
\end{equation}
which again one may prefer to re-express in terms of the lab-frame frequency
of the photon
\begin{equation}
\nu' \simeq \nu_*(1-v') - \frac{h \nu_*^2}{2m}~.
\end{equation}

So the lab-frame frequency difference between the two photons is
\begin{equation}
\Delta \nu =  \nu_* (v + v') + \frac{h \nu_*^2}{m} \simeq 2 v \nu_* + \frac{2 h \nu_*^2}{m}~,
\label{jocBOOSTIE1}
\end{equation}
and this (as easily seen upon noticing that in the nonrelativistic limit $v = p /m$)
of course perfectly agrees with the corresponding result
(\ref{deltaomeNOEP}), which we had obtained relying exclusively on
lab-frame kinematics.

It is easy to verify that redoing this Doppler-effect-based derivation
in presence of our Planck-scale corrections to the dispersion relation
(but setting aside, at least for now, possible Planck-scale dependence of
the Doppler effect) one ends up replacing (\ref{jocBOOSTIE1})
with
\begin{equation}
\Delta \nu =  \nu _* \left[ v(p) + v(p+ h \nu _* )\right]
+ \frac{ h \nu _*^2 }{m} + \xi_1 \frac{m}{M_P} \nu_*
~. \label{DeltaNUboostedGENERAL}
\end{equation}
This is the formula that should reproduce our main result (\ref{DeltaOmegaLeading}).
Indeed this is the point where one might encounter the necessity
of Planck-scale modifications of the boost/Doppler-effect
laws and/or of Planck-scale modifications
of the law that fixes the dependence of speeds on momentum/energy,
Concerning speeds if one assumes
 (as done by most
authors~\cite{grbgac,astroBiller,urrutiaPRL,PiranNeutriNat,LQGDispRel})
 $v = \partial E/\partial p$ then in our context (nonrelativistic regime, with $\xi_1$
parameter) one finds $v(p) = p/m + \xi_1 m/M_P$.
If instead, as argued by other authors~\cite{newVEL1,newVEL2,newVEL3},
consistency of the Planck-scale laws requires that $v = p/E$
should be enforced then in our
nonrelativistic context one of course has $v(p) = p/m$.

We find that the desirable agreement between (\ref{DeltaNUboostedGENERAL})
and (\ref{DeltaOmegaLeading})
is found upon assuming $v(p) = p/m$, which indeed allows one to rewrite
(\ref{DeltaNUboostedGENERAL}) as
\begin{equation}
\Delta \nu =   \frac{2 \nu_* (p + h \nu_*)}{m} + \xi_1 \frac{m}{M_P} \nu_*
~.
\end{equation}
If instead one insists on the alternative $v(p) = \partial E/\partial p = p/m + \xi_1 m/M_P$,
then (\ref{DeltaNUboostedGENERAL}) takes the form
\begin{equation}
\Delta \nu =   \frac{2 \nu_* (p + h \nu_*)}{m} + 2\xi_1 \frac{m}{M_P} \nu_*
~,
\end{equation}
which is sizably different from (\ref{DeltaOmegaLeading}).

Our observation that the law $v=p/m$ is automatically
consistent with a plausible symmetry-deformation perspective
is intriguing, but might well be just a quantitative accident.
We thought it might still be worth reporting just as a way to illustrate
the complexity of the issues that come into play if
our cold-atom studies are examined within a symmetry-deformation
scenario, issues that we postpone to future studies.
The Doppler effect in models with deformed Poincar\'e symmetries
had not been previously studied, and there are several alternative ``schools"
on how to derive from the energy-momentum dispersion relation a law
giving the speed as a function of energy.
In the specific case of the correction term we here parametrized with $\xi_1$
it would seem that $v=p/m$ is a natural choice, at least in as much as
the choice $v(p) = \partial E/\partial p$ appears to be rather pathological/paradoxical
since it leads to $v(p) = p/m + \xi_1 m$, {\it i.e.} a law
 that assigns nonzero speed to the particle even when
the spatial momentum vanishes.

\subsection{Testing energy-momentum conservation}\label{EPCONS}
Up to this point our analysis has focused on tests
of the Lorentz sector of Poincar\'e symmetry. But of course there is also interest
in testing the translation sector, and indeed there has been a corresponding effort,
particularly over the last decade.
The aspect of the translation sector on which these studies
have primarily focused is the law of energy-momentum conservation
in particle-physics processes, and
particularly noteworthy are some results~\cite{gactpPRD,sethEPCONS}
which exposed ``Planck-scale sensitivity" for the analysis of certain classes
of ``ultraviolet" (high-energy) modifications of the law of energy-momentum
conservation. Even for these studies one can contemplate
the alternative between breaking and deforming Poincar\'e symmetry, and from this perspective
it is rather noteworthy that the scenarios in which one deforms Poincar\'e symmetry
require~\cite{gacIJMP2002vD11,dsrIJMPrev} a consistency\footnote{These consistency requirements
for a
deformation of Poincar\'e symmetry are very restrictive but may not suffice to
fully specify the form of the law of energy-momentum conservation by insisting on
compatibility with a chosen form of the dispersion relation~\cite{gacIJMP2002vD11,dsrIJMPrev}.}
between the scheme of modification of the dispersion relation and
the scheme of modification of the law of energy-momentum conservation.
Instead of course if one is willing to break Poincar\'e symmetry one can consider
independently (or in combination) both
modifications of the dispersion relation and
modifications of the law of energy-momentum conservation.

In this Section we want to point out that our cold-atom-based strategy
also provides opportunities for studies of the form of the law of energy-momentum conservation
in the nonrelativistic regime. The observations on cold-atom experiments
that some of us reported in Ref.~\cite{gacPRL2009}
already inspired the recent analysis of Ref.~\cite{kowamich}, which provides
preliminary encouragement for the idea of using cold-atom experiments
for the study of the form of the law of energy-momentum conservation
in the nonrelativistic regime.
The scopes of the analysis reported in
Ref.~\cite{kowamich} were rather limited, since it focused on one specific model, which in
particular codifies no modifications of the dispersion relation: the only modification allowed
in Ref.~\cite{kowamich} appeared in the law of energy-momentum conservation and appeared
only at subleading order (in the sense here introduced in Sections~\ref{COMPLEMENTARITY}-\ref{nonrUR})
in the nonrelativistic limit.

While maintaining the perspective of a first exploratory investigation of these issues,
we shall here contemplate a more general scenario, with modifications of both
energy-momentum conservation and dispersion relation, and with correction
terms strong enough to appear even at the leading order in the nonrelativistic regime.
Besides aiming for greater generality, our interest in this direction is also motivated
by the desire of setting up future analysis which might consider in detail the interplay
between modifications of the dispersion relation and modifications of energy-momentum
conservation, particularly from the perspective of identifying scenarios
with deformation (rather than breakdown) of Poincar\'e symmetries, for which, as mentioned,
this interplay is in many instances required~\cite{gacIJMP2002vD11,dsrIJMPrev}.
While we shall not here attempt to formulate a suitable deformed-symmetry scenario,
the observations we here report are likely to be relevant for the possible future
search of such a formulation.

In light of the exploratory nature of our investigation of this point we shall be
satisfied illustrating the possible relevance of the interplay between dispersion relation
and energy-momentum conservation for the specific case of modified laws of
conservation of spatial momentum (ordinary conservation of energy):
\begin{equation}
\vec p_1 + \vec p_2 - \frac{\rho_1}{4 M_P} \left ( \frac{E_1^2}{E_2} \vec p_1
+ \frac{E_2^2}{E_1} \vec p_2 \right)  - \frac{\rho_2}{2 M_P}  ( E_1 \vec  p_2 + E_2  \vec  p_1)
=
\vec p_3 + \vec p_4 - \frac{\rho_1}{4 M_P}  \left ( \frac{E_3^2}{E_4} \vec p_3 + \frac{E_4^2}{E_3} \vec p_4 \right) - \frac{\rho_2}{2 M_P}  ( E_3 \vec  p_4 + E_4  \vec  p_3)
~.
\end{equation}
We are focusing on the case of
 two incoming and two outgoing particles (relevant for processes in which a photon is
 absorbed and one is emitted by an atom),
 and we characterized the modification in terms of parameters $\rho_1$ and  $\rho_2$.
As announced, we shall keep track of these parameters  $\rho_1$ and  $\rho_2$
together with the parameters $\xi_1$ and $\xi_2$ that parametrized the modifications
of the dispersion relation in the nonrelativsitic limit\footnote{Note that we have worked consistently
throughout this manuscript characterizing the nonrelativistic limit as the
long-wavelength
regime where $p \ll E$ for massive particles. This terminology was
inspired by our focus
on atoms and other massive particles, since the label ``nonrelativistic"
for long-wavelength
photons (massless particles) is of course not applicable. However, the
reader can easily check that
we handled correctly the long-wavelength properties of photons in the
relevant frameworks.
In particular, in our characterizations of
photons the parameter $\xi_1$ is automatically absent  (since $\xi_1$
always appears in formulas multiplied by powers of the mass of
the particle) and also parameters such as $\eta_1$ are omitted, since it
always appears
in the combination $\eta_1 p/M_P$ which in the long-wavelength
regime is completely negligible.}:
\begin{equation}
\left\{
\begin{array}{ll}
h \nu = h |\vec k|  +  \frac{\xi_2}{2 M_P} h^2 |\vec k|^2  & \text{(for photons)}
\\\\
E = m^2 + \frac{|\vec p|^2}{2m} +\frac{\xi_1}{2 M_P} m |\vec p|+  \frac{\xi_2}{2 M_P}  |\vec p|^2  &  \text{(for massive particles)}
\end{array}
\right.
\end{equation}

For a two-photon Raman transition our modified law of conservation
of spatial momentum has significant implications along the common direction
of the laser beams used to excite/de-excite the atoms:
\begin{eqnarray}
&&h |\vec k| + |\vec p| - \frac{\rho_1}{4 M_P}  \left ( \frac{h^2 \nu^2}{m}
h |\vec k| + \frac{m^2}{h \nu} |\vec p| \right)
-\frac{\rho_2}{2 M_P} ( h \nu |\vec  p| + E h |\vec  k|) = \nonumber \\
&&- h |\vec k'| + |\vec p'|
- \frac{\rho_1}{4 M_P}  \left ( - \frac{h^2 \nu'^2}{m} h |\vec k'| + \frac{m^2}{h \nu'} |\vec p'| \right)
- \frac{\rho_2}{2 M_P} ( h \nu' |\vec  p'| - E' h |\vec  k'|)  ~,
\label{jocnewlaw}
\end{eqnarray}

In Section~\ref{nonrUR} we used ordinary momentum conservation,
$h |\vec k| + |\vec p| = - h |\vec k'| + |\vec p'|$,
but if instead one adopts (\ref{jocnewlaw})
the following result is then straightforwardly obtained:
\begin{equation}
\frac{\Delta \nu}{2 \nu_*(\nu_* + p/h)} \simeq \frac{h}{m}
+ \frac{1}{M_P}\left[ m  (\xi_1 - \rho_1) + (2 \xi_2-\rho_2) p
+ 2 (\xi_2 - \rho_2) h \nu_* \right] \frac{h \nu_*}{2 \nu_*(h\nu_* + p)} ~.
\end{equation}
While this is, as stressed, only an exploratory investigation of the role that
could be played by modifications of energy-momentum conservation (in particular there
is clearly a strong influence of the specific {\it ansatz} we adopted for the
modified law of conservation of energy-momentum)
it is still noteworthy that the parameter $\rho_1$ enters the final result
at the same order as the parameter $\xi_1$
and similarly the parameter $\rho_2$ enters the final result
at the same order as the parameter $\xi_2$. In particular, this implies that even
at the type of leading nonrelativistic order we here mainly focused on (the order
where $\xi_1$ appears) the possibility of modifications of the law
of energy-momentum conservation may well be relevant, with nonnegligible effects
even in cases where $\xi_1 = 0$ but $\rho_1 \neq 0$.

\section{Closing remarks}
We have here used the noteworthy example of atom-recoil measurements
to explore whether it is possible to setup a phenomenology for the nonrelativistic
limit of the energy-momentum dispersion relation that adopts the same spirit
of a popular research program focusing instead on the corresponding ultrarelativistic
regime.
It appears that this is indeed possible and that on the one hand there is a strong complementarity
of insight to be gained by combining studies of the nonrelativistic regime  and of the
ultrarelativistic regime, and on the other hand the nature of the conceptual issues
that must be handled (particularly the relativistic issues associated with the
possibility of breaking or deforming Poincar\'e symmetry) are closely analogous.
We therefore argue that by adding the nonrelativistic limit to the relevant
phenomenology agenda we could improve our ability to
constrain certain scenarios, and we could also gain a powerful
tool from the conceptual side, exploiting the possibility to view the
same conceptual challenges within regimes that are otherwise very different.

In light of the remarkable pace of improvement of cold-atom experiments over the last 20 years,
we expect that the sensitivities here established
 for the parameters $\xi_1$ and $\xi_2$ (and $\lambda$; and $\rho_1$,$\rho_2$)
 might be significantly improved upon in the near
future. This will most likely translate into more stringent bounds, but,
particularly considering the values of $\xi_1$ being probed,
should also be viewed as a (slim but valuable) chance for
a striking discovery. We therefore feel
that our analysis should motivate
experimentalists to taylor some of their plans in this direction
(also using the remarks we offered in Subsection~\ref{NEWLIMIT})
and should motivate theorists toward
a vigorous effort  aimed
at overcoming the technical difficulties on the quantum-gravity-theory side
that presently
obstruct the derivation of more detailed quantitative predictions.

\section*{Acknowledgments}
G.~A.-C. is supported in part by grant RFP2-08-02 from
The Foundational Questions Institute (fqxi.org). C. L. is
supported in part by the German Research Foundation and
the Centre for Quantum Engineering and Space-Time
Research QUEST.
D. M., J. M. C., J. L. C. and J. I. are supported by CICYT
(grants FPA2006-02315 and FPA2009-09638) and DGIID-DGA (grant
2009-E24/2). J. I. acknowledges a FPU grant and D. M. a FPI grant from
MICINN.


\begin{thebibliography}{50}

\bibitem{gacLRR}
G.~Amelino-Camelia,
gr-qc/9910089,
Lect.~Notes Phys.~ \textbf{541}, 1 (2000);
{\it Quantum Gravity Phenomenology}, arXiv:0806.0339.

\bibitem{grbgac}
G. Amelino-Camelia \emph{et~al.},
astro-ph/9712103,
Nature \textbf{393}, 763
(1998).

\bibitem{astroSchaefer}
B.E.~Schaefer, Phys. Rev Lett. \textbf{82}, 4964
(1999).

\bibitem{astroBiller}
S.D. Biller \emph{et~al.},
Phys. Rev. Lett. \textbf{83}, 2108
(1999).

\bibitem{gacNature1999}
G. Amelino-Camelia,
gr-qc/9808029, Nature {\bf 398}, 216
(1999).

\bibitem{urrutiaPRL}
J. Alfaro, H. A. Morales-Tecotl and L. F. Urrutia,
Phys. Rev. Lett. \textbf{84}, 2318
(2000).

\bibitem{jaconature}
T. Jacobson, S. Liberati and D. Mattingly,
astro-ph/0212190,
Nature {\bf 424}, 1019
(2003).


\bibitem{Gaclaem}
G. Amelino-Camelia and C. L\"ammerzahl,
gr-qc/0306019,
Class. Quant. Grav. \textbf{21}, 899
(2004).

\bibitem{PiranNeutriNat}
U. Jacob and T. Piran,
hep-ph/0607145,
Nature Phys. {\bf 3}, 87
(2007).

\bibitem{hessPRL}
F. Aharonian et al. [HESS Collaboration], 
Phys. Rev. Lett. {\bf 101}, 170402 (2008).

\bibitem{gacPRL2009}
G.~Amelino-Camelia, C.~Laemmerzahl, F.~Mercati and G.~M.~Tino,
arXiv:0911.1020 [gr-qc], Phys.\ Rev.\ Lett.\  {\bf 103}, 171302 (2009).

\bibitem{LQGDispRel}
R. Gambini and J. Pullin,
Phys. Rev. \textbf{D59}, 124021 (1999).

\bibitem{gacmajid} G. Amelino-Camelia and S. Majid,
hep-th/9907110, Int. J. Mod. Phys. \textbf{A15}, 4301
(2000).

\bibitem{kowaPLBcosmo} J. Kowalski-Glikman,
astro-ph/0006250, Phys.~Lett.~\textbf{B499}, 1 (2001).

\bibitem{Orfeupion}
O. Bertolami and L. Guisado,
hep-th/0306176, JHEP {\bf 0312}, 013 (2003).

\bibitem{smolinbook}
L. Smolin,
\emph{Three roads to quantum gravity}
(Basic Books, 2002).

\bibitem{astroKifune}
T. Kifune,
astro-ph/9904164, Astrophys. J. Lett. \textbf{518}, L21
(1999).

\bibitem{gacQM100}
G. Amelino-Camelia,
gr-qc/0012049, Nature {\bf 408}, 661
(2000).


\bibitem{wilcGUTEP}
S. P. Robinson and F. Wilczek,
hep-th/0509050, Phys. Rev. Lett. {\bf 96}, 231601
(2006).

\bibitem{fermiSCIENCE}
A.~Abdo \emph{et al.},
Science {\bf 323}, 1688
(2009).

\bibitem{ellisPLB2009}
J.~Ellis, N.~E.~Mavromatos and D.~V.~Nanopoulos,
arXiv:0901.4052 [astro-ph], Phys.\ Lett.\  B {\bf 674}, 83 (2009).
  
\bibitem{gacSMOLINprd2009}
G.~Amelino-Camelia and L.~Smolin,
arXiv:0906.3731 [astro-ph], Phys.\ Rev.\  D {\bf 80}, 084017 (2009).

\bibitem{fermiNATURE}
A.~Abdo \emph{et al.},
Nature \textbf{462}, 331 (2009).

\bibitem{gacNATURE2009}
G. Amelino-Camelia,
Nature \textbf{462}, 291 (2009).

\bibitem{urrutiaPRD}
J.~Alfaro, H.A.~Morales-Tecotl, L.F.~Urrutia,
hep-th/0208192, Phys.~Rev.~{\bf D66}, 124006 (2002).

\bibitem{gacIJMP2002vD11}
G.~Amelino-Camelia,
gr-qc/0012051, Int.~J.~Mod.~Phys.~{\bf D11}, 35 (2002).

\bibitem{gacdsrPLB2001}
G.~Amelino-Camelia,
hep-th/0012238, Phys.\ Lett.\  B {\bf 510}, 255 (2001).

\bibitem{leedsrPRL}
J.~Magueijo and L.~Smolin,
Phys.\ Rev.\ Lett.\  {\bf 88}, 190403
(2002).

\bibitem{Kasevich91b}
M. Kasevich and S. Chu,
Phys. Rev. Lett. \textbf{67}, 181 (1991).

\bibitem{Peters99}
A. Peters \emph{et~al.},
Nature \textbf{400}, 849
(1999).

\bibitem{Wicht02}
A. Wicht \emph{et al.},
Phys. Script. \textbf{T102}, 82 (2002).

\bibitem{gab08}
D. Hanneke, S. Fogwell and G. Gabrielse,
Phys. Rev. Lett. \textbf{100}, 120801 (2008).

\bibitem{Gerginov06}
V. Gerginov \emph{et al.},
Phys. Rev. {\bf A73}, 032504 (2006).

\bibitem{liberatiNONUNIV}
S.~Liberati and L.~Maccione,
Ann.\ Rev.\ Nucl.\ Part.\ Sci.\  {\bf 59}, 245 (2009).

\bibitem{dsrIJMPrev}
G.~Amelino-Camelia,
 gr-qc/0210063, Int.~J.~Mod.~Phys.~{\bf D11}, 1643 (2002).
 
\bibitem{HIN}
F. Hinterleitner,
gr-qc/0409087, Phys. Rev. D {\bf 71}, 025016 (2005).

\bibitem{HOS}
S. Hossenfelder,
hep-th/0702016, Phys. Rev. D {\bf 75}, 105005 (2007).

\bibitem{ROV}
C. Rovelli,
arXiv:0808.3505 [gr-qc].

\bibitem{LIV-tritium}
J.M. Carmona, J.L. Cort\'es,
hep-ph/0007057, Phys. Lett. B {\bf 494}, 75 (2000).

\bibitem{LIV-cutoffs}
J.M. Carmona, J.L. Cort\'es,
hep-th/0012028, Phys. Rev. D {\bf 65}, 025006 (2001).

\bibitem{ellisPHOTONonly}
J. Ellis, N.E. Mavromatos, D.V. Nanopoulos, A.S. Sakharov,
astro-ph/0309144v2, Nature {\bf 428}, 386 (2004).

\bibitem{udem}
T. Udem,
Nature Phys. \textbf{2}, 153
(2006).

\bibitem{biraben06}
P. Clad\'e \textit{et al.},
Phys. Rev. A \textbf{74}, 052109 (2006).

\bibitem{biraben08}
M. Cadoret \emph{et~al.},
Phys. Rev. Lett. \textbf{101}, 23080 (2008).

\bibitem{birabenPROCEEDINGS}
F.~Biraben,
\emph{Proceedings of the XXI International Conference on Atomic Physics},
56 (World Scientific, 2009).

\bibitem{gacflavtsviuri}
U.~Jacob, F.~Mercati,  G.~Amelino-Camelia and T.~Piran, arXiv:1004.0575 [astro-ph].

\bibitem{newVEL1}
P.~Kosinski \& P.~Maslanka,
hep-th/0211057, Phys.~Rev.~D {\bf 68}, 067702 (2003).

\bibitem{newVEL2}
S.~Mignemi,
hep-th/0302065,
Phys.~Lett.~A {\bf316}, 173 (2003).

\bibitem{newVEL3}
M.~Daszkiewicz, K.~Imilkowska \& J.~Kowalski-Glikman,
hep-th/0304027,
Phys.Lett. A {\bf 323}, 345(2004).

\bibitem{gactpPRD}
G.~Amelino-Camelia \& T.~Piran,
astro-ph/0008107,
Phys.~Rev.~D {\bf 64},  036005 (2001).

\bibitem{sethEPCONS}
D.~Heyman, F.~Hinteleitner \& S.~Major,
gr-qc/0312089,
Phys.~Rev.~D {\bf 69}, 105016 (2004).

\bibitem{kowamich}
M. Arzano, J. Kowalski-Glikman, A. Walkus,
arXiv:0912.2712 [hep-th].

\end{thebibliography}
\end{document}